\documentclass[journal]{IEEEtran}
\usepackage[pdftex,dvipdfmx]{graphicx}
\usepackage[subrefformat=parens]{subcaption}
\usepackage{textcomp}
\usepackage[noadjust]{cite}
\usepackage[T1]{fontenc} 
\usepackage{amsmath}
\usepackage[cmintegrals]{newtxmath}
\usepackage{bm}
\usepackage{array}

\begin{document}
\title{Complex Number Assignment in the Topology Method for Heartbeat Interval Estimation Using Millimeter-Wave Radar}

\author{Yuji~Tanaka,~\IEEEmembership{Member, IEEE}, Kimitaka~Sumi, Itsuki~Iwata,~\IEEEmembership{Graduate Student Member, IEEE} and Takuya~Sakamoto,~\IEEEmembership{Senior Member, IEEE}
  \thanks{Y.~Tanaka, K.~Sumi, S.~Iwata and T.~Sakamoto are with the Department of Electrical Engineering, Graduate School of Engineering, Kyoto University, Kyoto 615-8510, Japan.}
}
\markboth{}%
{Tanaka \emph{et al.}: Complex Number Assignment in the Topology Method for Heartbeat Interval Estimation Using Millimeter-Wave Radar}

\maketitle
\begin{abstract}
The topology method is an algorithm for accurate estimation of instantaneous heartbeat intervals using millimeter-wave radar signals. In this model, feature points are extracted from the skin displacement waveforms generated by heartbeats and a complex number is assigned to each feature point. However, these numbers have been assigned empirically and without solid justification.
This study used a simplified model of displacement waveforms to predict the optimal choice of the complex number assignments to feature points corresponding to inflection points, and the validity of these numbers was confirmed using analysis of a publicly available dataset. 
\end{abstract}


\IEEEpeerreviewmaketitle

\section{Introduction}
Global population aging creates a high demand for continuous monitoring of physiological signals~\cite{bib:Chen2023}.
Among these, heart rate is particularly important for monitoring health status and detecting signs of cardiovascular disease.
In addition, heart rate variability  can be used to capture emotional changes~\cite{bib:Quintana2012, bib:Zhao2018} and monitor mental state~\cite{bib:Han2019}.
The most common methods for monitoring heart rate are electrocardiography and photoplethysmography. 
These use contact-type sensors that can cause discomfort and restrict the wearer’s activities~\cite{bib:Kitagawa2022} and are therefore unsuitable for long-term monitoring applications.
In contrast, radar-based methods that measure body surface displacement caused by heart motion and pulse waves have attracted attention because they enable long-term heartbeat monitoring without sensor attachment~\cite{bib:Nosrati2018, bib:Petrovic2019, bib:Ye2020, bib:Wang2021, bib:Chen2022}.

The topology method~\cite{bib:Sakamoto2016} is an accurate technique for estimating the instantaneous heartbeat interval using radar signals and has found various applications such as autonomic nervous system activity estimation~\cite{bib:Sakamoto2020}, camera-based heart rate estimation~\cite{bib:Sakamoto2022}, and animal monitoring~\cite{bib:Iwata2023}.
In the topology method, several feature points are extracted from the skin displacement waveform caused by heartbeats, and a complex number is assigned to each of the feature points.
However, studies to date have set these numbers empirically, and thus the question of how to validate the complex number assignments remains open.
In this study, we used a simplified model of the displacement waveform to quantitatively evaluate the validity of the complex numbers assigned to feature points corresponding to two types of inflection points, and we analyzed its performance.

\section{Topology Method}
The topology method~\cite{bib:Sakamoto2016} can accurately estimate the interbeat interval (IBI) of the heart by using topology correlation coefficients calculated from the complex numbers assigned to the extracted feature points in combination with ordinary correlation coefficients.
As shown in Table~\ref{tbl:ComparisonTable}, it is reported that the topology method achieves high accuracy in estimating IBI in comparison with other representative studies~\cite{bib:Wang2021, bib:Chen2022}.
The topology method extracts the following six types of feature points from the displacement waveform~$s(t)$:
\begin {itemize}
	\item PK (peak): $\dot{s}(t) = 0$, $\ddot{s}(t) < 0$,
	\item VL (valley): $\dot{s}(t) = 0$, $\ddot{s}(t) > 0$,
	\item RDP (rising derivative peak): $\dot{s}(t) > 0$, $\ddot{s}(t) = 0$, $\dddot{s}(t) < 0$,
	\item RDV (rising derivative valley):\\ $\dot{s}(t) > 0$, $\ddot{s}(t) = 0$, $\dddot{s}(t) > 0$,
	\item FDP (falling derivative peak): $\dot{s}(t) < 0$, $\ddot{s}(t) = 0$, $\dddot{s}(t) < 0$,
	\item FDV (falling derivative valley):\\ $\dot{s}(t) < 0$, $\ddot{s}(t) = 0$, $\dddot{s}(t) > 0$.
\end{itemize} 
To estimate the IBI, the topology method detects pairs of feature points of the same type that have high topology and ordinary correlation coefficients. 

\begin{table}[tb]
\caption{Comparison of the interbeat interval estimation accuracy between the topology method and other methods.} 
\begin{tabular}{c||c|c|c|}
 &  Wang et al.~\cite{bib:Wang2021} & Chen et al.~\cite{bib:Chen2022} & Iwata et al.~\cite{bib:Iwata2023} \\
 \hline \hline
Method &  VMD & DNN & Topology + HPF \\
Radar type & FMCW & FMCW & FMCW \\
Center frequency & 79 GHz & 79 GHz & 79 GHz \\
RMS error & 26 ms & 3 ms & 2.55 ms\\
\hline
\end{tabular}
VMD: variational mode decomposition, DNN: deep neural network,\\
HPF: high-pass filter, FMCW: frequency-modulated continuous-wave,\\
RMS: root mean square
\label{tbl:ComparisonTable}
\end{table}

First, the ordinary correlation coefficients are calculated.
Let $\tau_n$ be the time of the $n$th feature point extracted from $s(t)$.
The $(2K + 1)$-dimensional vector, $\overline{\boldsymbol{v}}_n$, composed of the signal samples around the $n$th feature point is expressed as $\overline{\boldsymbol{v}}_n = \left[s\left(\tau_n-K \Delta t\right), s\left(\tau_n-(K-1) \Delta t\right), \ldots, s\left(\tau_n+K \Delta t\right)\right]^{\mathsf{T}}$,
where $\Delta t$ is the sampling interval, $K$ is a parameter determined by $K \Delta t = T_{\mathrm{c}}/2$, $T_{\mathrm{c}}$ is the time width used to calculate an ordinary correlation coefficient, and the superscript $\mathsf{T}$ denotes the matrix transpose.
Let $\boldsymbol{v}_n$ be the vector obtained by removing the DC component from $\overline{\boldsymbol{v}}_n$.
The ordinary correlation between $m$th and $n$th feature points is then given by the coefficient
$c_{m, n}={\boldsymbol{v}_m^{\mathsf{T}} \boldsymbol{v}_n}/{\left|\boldsymbol{v}_m\right|\left|\boldsymbol{v}_n\right|}$.

\begin{figure}[tb]
  \begin{minipage}[H]{0.48\columnwidth}
    \centering
    \includegraphics[width=\columnwidth]{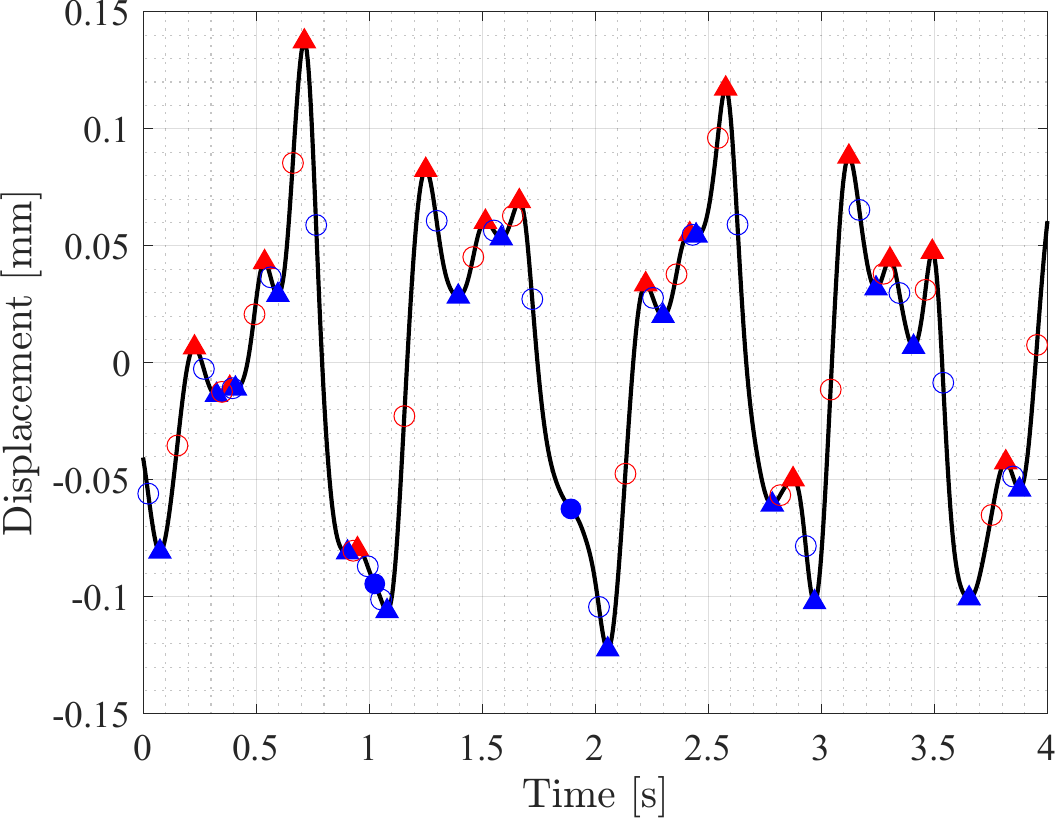}
    \subcaption{Example of the displacement waveform $s(t)$}
    \label{fig:Radar-example}
  \end{minipage}
  \hspace{0.04\columnwidth} 
  \begin{minipage}[H]{0.48\columnwidth}
    \centering
    \includegraphics[width=\columnwidth]{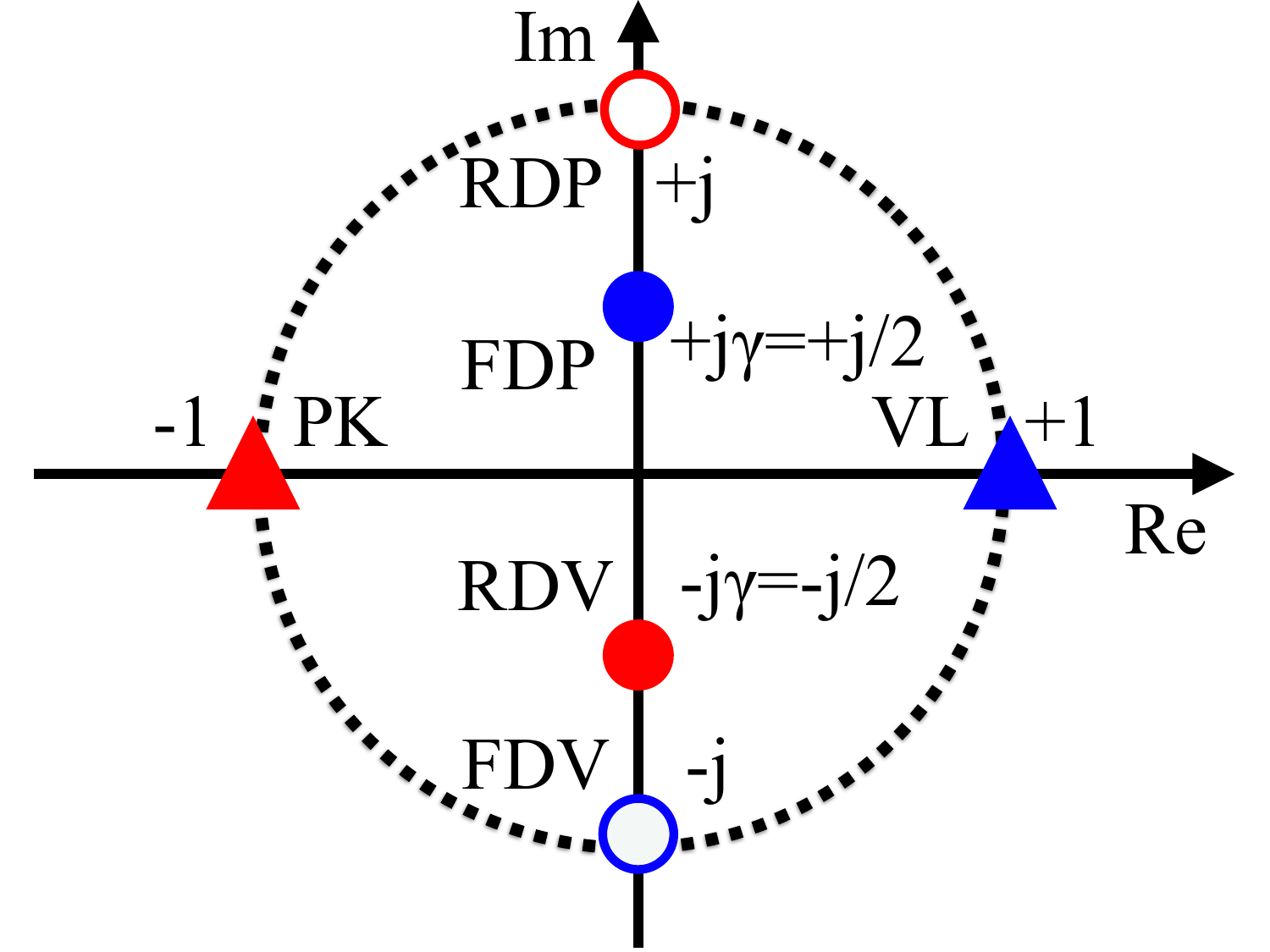}
    \subcaption{Feature points}
    \label{fig:TopologyNum_0}
  \end{minipage}
  \caption{Complex values assigned to feature points from which the topology correlation values are calculated.} 
  \label{fig:TopologyNum}
\end{figure}

Next, the topology correlation coefficients are calculated.
In the topology method, the complex numbers $-1$, $1$, $\mathrm{j}$, $-\mathrm{j}\gamma$, $\mathrm{j}\gamma$, and $-\mathrm{j}$ are assigned to the feature points PK, VL, RDP, RDV, FDP, and FDV, respectively, where $\gamma = 1/2$ and $\mathrm{j} = \sqrt{-1}$ is an imaginary unit (Fig.~\ref{fig:TopologyNum}).
Each symbol shown in Fig.~\ref{fig:TopologyNum}(a) corresponds to the same symbol shown in Fig.~\ref{fig:TopologyNum}(b).
Based on these assignments, the complex signal $s_{\mathrm{t}}(t)$ is calculated by referring to the nearest feature point at time $t$.
To obtain the topology correlation coefficient from $s_{\mathrm{t}}(t)$, define a $(2K_{\mathrm{t}} + 1)$-dimensional complex vector $\boldsymbol{u}_n$ for the $n$th feature point as
$\boldsymbol{u}_n=\left[s_{\mathrm{t}}\left(\tau_n-K_{\mathrm{t}} \Delta t\right), s_{\mathrm{t}}\left(\tau_n-\left(K_{\mathrm{t}}-1\right) \Delta t\right), \ldots, s_{\mathrm{t}}\left(\tau_n+K_{\mathrm{t}} \Delta t\right)\right]^\mathsf{T}$,
where $K_{\mathrm{t}}$ is a parameter determined by $K_{\mathrm{t}} \Delta t = T_{\mathrm{t}}/2$, and $T_{\mathrm{t}}$ is the time width used to calculate the topology correlation values.
The topology correlation $q_{m,n}$ between the $m$th and $n$th feature points is expressed as
$q_{m, n}={\left|\boldsymbol{u}_m^{\mathsf{H}} \boldsymbol{u}_n\right|^2}/{\left|\boldsymbol{u}_m\right|^2\left|\boldsymbol{u}_n\right|^2}$,
where the superscript $\mathsf{H}$ denotes the complex conjugate of the matrix transpose.
The set of feature points that satisfy both $c_{m, n} \geq c_{\mathrm{th}}$ and $q_{m, n} \geq q_{\mathrm{th}}$ are extracted, where $c_{\mathrm{th}}$ and $q_{\mathrm{th}}$ are thresholds for the ordinary correlation and the topology correlation coefficients, respectively.
Finally, the IBI is estimated from the time difference between two adjacent feature points of the same type. 

Note that the parameter $\gamma = 0.5$ is empirically set, so RDV and FDP, the feature points corresponding to inflection points, are empirically assigned the values $-\mathrm{j}/2$ and $\mathrm{j}/2$, respectively.
Thus, the following factors should be considered.
The first is the magnitude relationship between RDV and FDP.
Considering that $\dot{s}(t) > 0$ for RDV and $\dot{s}(t) < 0$ for FDP, the number assigned to the imaginary part of RDV appears to exceed that assigned to FDP, but the actual relationship is opposite.
In the original formulation of the topology method~\cite{bib:Sakamoto2016} this opposite assignment was shown to make the peak of the cross-correlation function steeper, but the effectiveness of such an assignment has not been verified quantitatively.
The second factor is the magnitude of the assigned numbers.
In the topology method, $\ddot{s}(t)$ and $\dot{s}(t)$ correspond to the real and imaginary parts of the complex number assignment, respectively.
The only restriction imposed by the topology method on RDV and FDP, both of which satisfy $\ddot{s}(t) = 0$, is that the assigned numbers must be pure imaginary.
The magnitudes of the coefficients for RDV and FDP do not necessarily satisfy $|\gamma| = 1/2$.
To improve the accuracy of the topology method for estimating IBI, it is necessary to quantitatively analyze the effect of the above parameter assignments on method performance.

\section{Evaluation of Assigned Numbers Using a Simplified Model}
In this study, we used a simplified model of skin displacement to evaluate the effect of the numbers assigned to RDV and FDP on the performance of the topology method.
The simplified model of body displacement $d(t)$ measured using a radar system is expressed as
$d(t) = d_0 + d_{\mathrm{T}}(t) + d_{\mathrm{R}}(t) + d_{\mathrm{H}}(t)$,
where $d_0$ is the mean distance to the reflection point and $d_{\mathrm{T}}(t)$, $d_{\mathrm{R}}(t)$, and $d_{\mathrm{H}}(t)$ are the displacements associated with body motion, respiration, and heartbeat, respectively.
We assumed that $d_{\mathrm{T}}(t)$, $d_{\mathrm{R}}(t)$, and $d_{\mathrm{H}}(t)$ are all periodic functions in the local time range and have different fundamental frequencies.
We also assumed that harmonics of $d_{\mathrm{H}}(t)$ higher than 3rd order are negligible.
In this case, the displacement waveform $s(t)$, obtained using an ideal bandpass filter, consists only of the fundamental wave and 2nd harmonic of the heartbeat displacement:
\begin{align}
s(t) = d_{\mathrm{H}}(t) 
:= \mathrm{cos} \left( \omega_0 t \right) + \alpha \mathrm{cos} \left( 2 \omega_0 t + \theta \right),
\label{eq:signal_heart_analyze}
\end{align}
where $\omega_0$, $\alpha$ and $\theta$ are the angular frequency of the fundamental wave, the amplitude and the initial phase of the 2nd harmonic, respectively.
We evaluated the assigned numbers with respect to the fundamental wave, so the amplitude and the initial phase of the fundamental wave can be set to 1 and 0, respectively.
The amplitude of the 2nd harmonic of the heartbeat component is often smaller than that of the fundamental wave~\cite{bib:Nguyen2013, bib:Xiong2020}, so $0 \leq \alpha < 1$ is satisfied.

The six types of feature points (PK, VL, RDP, RDV, FDP, and FDV) are determined from the 1st-, 2nd-, and 3rd-order derivatives of the displacement waveform $s(t)$.
The 1st-order derivative is expressed as
\begin{align}
\dot{s}(t) 
= \beta_1 \mathrm{sin} \left( \omega_0 t \right) + \beta_2  \mathrm{sin} \left( 2 \omega_0 t + \theta \right),
\label{eq:signal_1st_derivative}
\end{align}
where $\beta_1 = -\omega_0$ and $\beta_2 = -2\omega_0 \alpha$.
Similar to $s(t)$, $\dot{s}(t)$ is composed of the fundamental wave and 2nd~order harmonic. 
Similarly, the 2nd- and 3rd-order derivatives are expressed as follows:
\begin{align}
&\ddot{s}(t) 
= \omega_0 \left[ \beta_1 \mathrm{cos} \left( \omega_0 t \right) + 2 \beta_2  \mathrm{cos} \left( 2 \omega_0 t + \theta \right) \right],
\label{eq:signal_2nd_derivative}
\\
&\dddot{s}(t) 
= -\omega_0^2 \left[ \beta_1 \mathrm{sin} \left( \omega_0 t \right) + 4 \beta_2  \mathrm{sin} \left( 2 \omega_0 t + \theta \right) \right].
\label{eq:signal_3rd_derivative}
\end{align}

Consider the case where $s(t)$ is represented by a fundamental wave (i.e., $\alpha = 0$).
Then
$\dddot{s}(t)
= -\omega_0^2 \beta_1 \mathrm{sin} \left( \omega_0 t \right)
= -\omega_0^2 \cdot \dot{s}(t)$,
so that $\dot{s}(t)$ and $\dddot{s}(t)$ have different signs.
This indicates that among the four types of feature points satisfying $\ddot{s}(t) = 0$, only RDP and FDV appear when $s(t)$ is a fundamental wave.
In other words, RDV and FDP appear as feature points only when $s(t)$ contains harmonics.

Next, consider the conditions under which the feature point RDV appears.
Since $\dot{s}(t) > 0$, $\ddot{s}(t) = 0$, and $\dddot{s}(t) > 0$ are satisfied for RDV, the following are obtained using Eqs.~\eqref{eq:signal_1st_derivative},~\eqref{eq:signal_2nd_derivative} and \eqref{eq:signal_3rd_derivative}: 
\begin{align}
&\beta_1 \mathrm{sin} \left( \omega_0 t \right) + \beta_2  \mathrm{sin} \left( 2 \omega_0 t + \theta \right) > 0,
\label{eq:RDV_condition_1st_derivative}
\\
&\beta_1 \mathrm{cos} \left( \omega_0 t \right) + 2 \beta_2  \mathrm{cos} \left( 2 \omega_0 t + \theta \right) = 0,
\label{eq:RDV_condition_2nd_derivative}
\\
&\beta_1 \mathrm{sin} \left( \omega_0 t \right) + 4 \beta_2  \mathrm{sin} \left( 2 \omega_0 t + \theta \right) < 0.
\label{eq:RDV_condition_3rd_derivative}
\end{align}
Using Eqs. \eqref{eq:RDV_condition_1st_derivative} and \eqref{eq:RDV_condition_3rd_derivative}, we obtain
\begin{align}
\beta_1 \mathrm{sin} \left( \omega_0 t \right) > 0, \quad \beta_2  \mathrm{sin} \left( 2 \omega_0 t + \theta \right) < 0.
\label{eq:basic-harmonics_diff-sign}
\end{align}
From Eq.~\eqref{eq:signal_1st_derivative}, we can confirm that the fundamental and harmonic components of $\dot{s}(t)$ have different signs when RDV feature points appear.
When $\alpha = 0$, $\dot{s}(t) = \beta_1 \mathrm{sin} \left(\omega_0 t \right) > 0$ holds, so RDP and RDV can be compared.
The RDP is a feature point that mainly corresponds to the fundamental component, whereas the RDV is a feature point that corresponds to the harmonic component.
Therefore, from Eq.~\eqref{eq:basic-harmonics_diff-sign}, it is reasonable to allow the coefficients assigned to RDP and RDV to have different signs when both feature points corresponding to the fundamental component and the harmonic component are evaluated using topology correlation coefficients. It can be seen that the coefficients assigned to RDP and RDV should have opposite signs (i.e. $\gamma > 0$ should be satisfied).

In addition, from Eq. \eqref{eq:RDV_condition_1st_derivative}, $\beta_1^2 \mathrm{sin}^2 \left( \omega_0 t \right) > \beta_2^2  \mathrm{sin}^2 \left( 2 \omega_0 t + \theta \right)$ is obtained, and thus
\begin{align}
\rho^2 < \frac{4 - 3\mathrm{cos}^2 \left( \omega_0 t \right)}{4}, \quad \rho = \left|\frac{\beta_2}{\beta_1}\right| = 2 \alpha \left( < 2 \right)
\label{eq:rho-square_upper-bound}
\end{align}
is obtained from Eq.~\eqref{eq:RDV_condition_2nd_derivative}.
Note that $\rho$ and $\gamma$ correspond to the coefficient ratio of the fundamental frequency and harmonics of $\dot{s}(t)$, where $\rho$ considers only the absolute values, whereas $\gamma$ considers the values with signs.
Similarly, from Eqs. \eqref{eq:RDV_condition_1st_derivative} and \eqref{eq:RDV_condition_3rd_derivative}, the following inequality is obtained:
\begin{align}
\frac{1 + 3\mathrm{cos}^2 \left( \omega_0 t \right)}{16} < \rho^2.
\label{eq:rho-square_lower-bound}
\end{align}
Assume that $\rho^2$ is uniformly distributed in Eqs.~\eqref{eq:rho-square_upper-bound} and \eqref{eq:rho-square_lower-bound}.
The time average of $\rho^2$, denoted as $\bar{\rho}^2$, is expressed as
\begin{align}
\bar{\rho}^2 
= \frac{1}{\pi} \int_{\pi}^{2\pi} \frac{1}{2} \left( \frac{4 - 3\mathrm{cos}^2 \left( x \right)}{4} + \frac{1 + 3\mathrm{cos}^2 \left( x \right)}{16} \right) ~\mathrm{d}x
= \left( \frac{5}{8} \right)^2,
\end{align}
where $x = \omega_0 t$ and $\pi \leq x < 2\pi$ satisfies because of the first inequality of Eq.~\eqref{eq:basic-harmonics_diff-sign}.
Considering that $\gamma > 0$, the suitable number assignment for RDV is expected to be $-\mathrm{j}5/8$.
In this case, the suitable number assignment for FDP is then $\mathrm{j}5/8$ because FDP is the negative of RDV.
Therefore, we can conclude that the optimal coefficient for RDV and FDP satisfies $\gamma = 5/8 = 0.625$, which is larger than $\gamma = 1/2$ used in the original study~\cite{bib:Sakamoto2016}.

\section{Analysis by Dataset}
\subsection{Evaluation Specification}
We evaluated the accuracy of the IBI estimate by changing the coefficient $\gamma$ in the topology method and using a publicly available dataset provided by Schellenberger~et al.~\cite{bib:Schellenberger2020}. 
To calculate the ordinary and topology correlation coefficients, we set the window length as $T_{\mathrm{c}} = T_{\mathrm{t}} = 0.5$~s.
The average overlapped length is 0.33 s, so the average overlap ratio is 66\%. Note that the overlapped length depends on the intervals between feature points.
The thresholds $c_{\mathrm{th}}$ and $q_{\mathrm{th}}$ for the ordinary and topology correlation coefficients are set to $c_{\mathrm{th}} = 0.7$ and $q_{\mathrm{th}} = 0.5$, respectively.
We also assumed that the IBI lies within the 0.4\textendash1.2~s range, which corresponds to heart rates of 50\textendash150 bpm.
Complex numbers $-\mathrm{j}\gamma$ and $\mathrm{j}\gamma$ are assigned to feature points RDV and FDP, respectively, where the parameter $\gamma$ is set to $\gamma = -2, -1.875, -1.75, \cdots, 1.75, 1.875$, or $2$. 

For the evaluation, we use the dataset provided by Schellenberger~et al.~\cite{bib:Schellenberger2020}.
This dataset consists of 10-minute measurement data for each person obtained from 30 participants at rest using a 24 GHz continuous-wave radar system with a sampling frequency of 2 kHz.
We resolve the DC offset by subtracting the time average value of the IQ plot. 
We calculate the phase using four-quadrant inverse tangent and unwrap the phase when the phase gap between consecutive samples is greater than or equal to $\pi$, and then $M \times 2\pi$ ($M$ is a nonzero integer) is added so that the phase gap becomes less than $\pi$.
To extract the heartbeat component using the topology method, a high-pass finite impulse response filter with a cutoff frequency of $0.5$ Hz and a stopband attenuation of 60 dB was applied to the displacement waveform derived from the phase component of the radar signal.
Note that the third harmonic of the respiration signal might affect the accuracy in estimating IBI, which needs to be addressed in future works.

The accuracy of the topology method was evaluated using the root mean square (RMS) error between the estimated IBI and a reference value obtained from an electrocardiograph (ECG) synchronized with the radar.
In this study, we evaluated data pertaining to 21 of the 30 participants. Participant numbers 1\textendash 4 and 25 were excluded because no ECG data were recorded, and numbers 10, 16, 19, and 30 were excluded because their IBIs were outside the assumed 0.4\textendash1.2 s range.
We also used the time coverage rate (TCR) \cite{bib:Sakamoto2021} as another evaluation criterion.
The TCR is defined as $k_\varepsilon \Delta t_{\mathrm{TCR}} / T_{\mathrm{all}}$, where $k_\varepsilon$ is the number of intervals that contain at least one accurately estimated point with an error of less than $\varepsilon$, $\Delta t_{\mathrm{TCR}}$ is the length of each time interval, and $T_{\mathrm{all}}$ is the total measurement time.
Following Sakamoto~et al.~\cite{bib:Sakamoto2021}, we set $\varepsilon = 50$ ms and $\Delta t_{\mathrm{TCR}} = 1.0$ s.

\subsection{Performance Evaluation}
\begin{figure*}[tb]
  \begin{minipage}[H]{0.32\linewidth}
    \centering
    \includegraphics[width=\columnwidth]{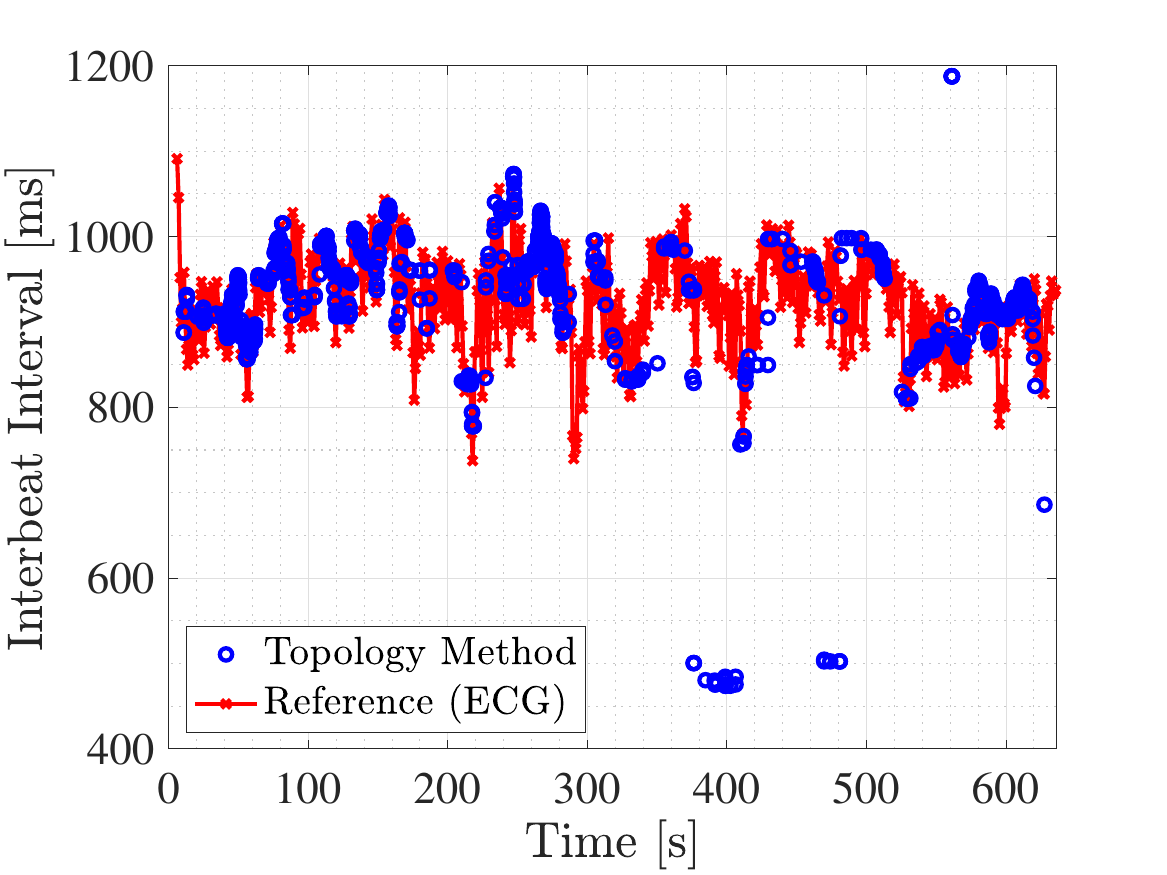}
    \subcaption{$\gamma = 0$}
    \label{fig:pm0}
  \end{minipage}
  \hspace{0.02\columnwidth} 
  \begin{minipage}[H]{0.32\linewidth}
    \centering
    \includegraphics[width=\columnwidth]{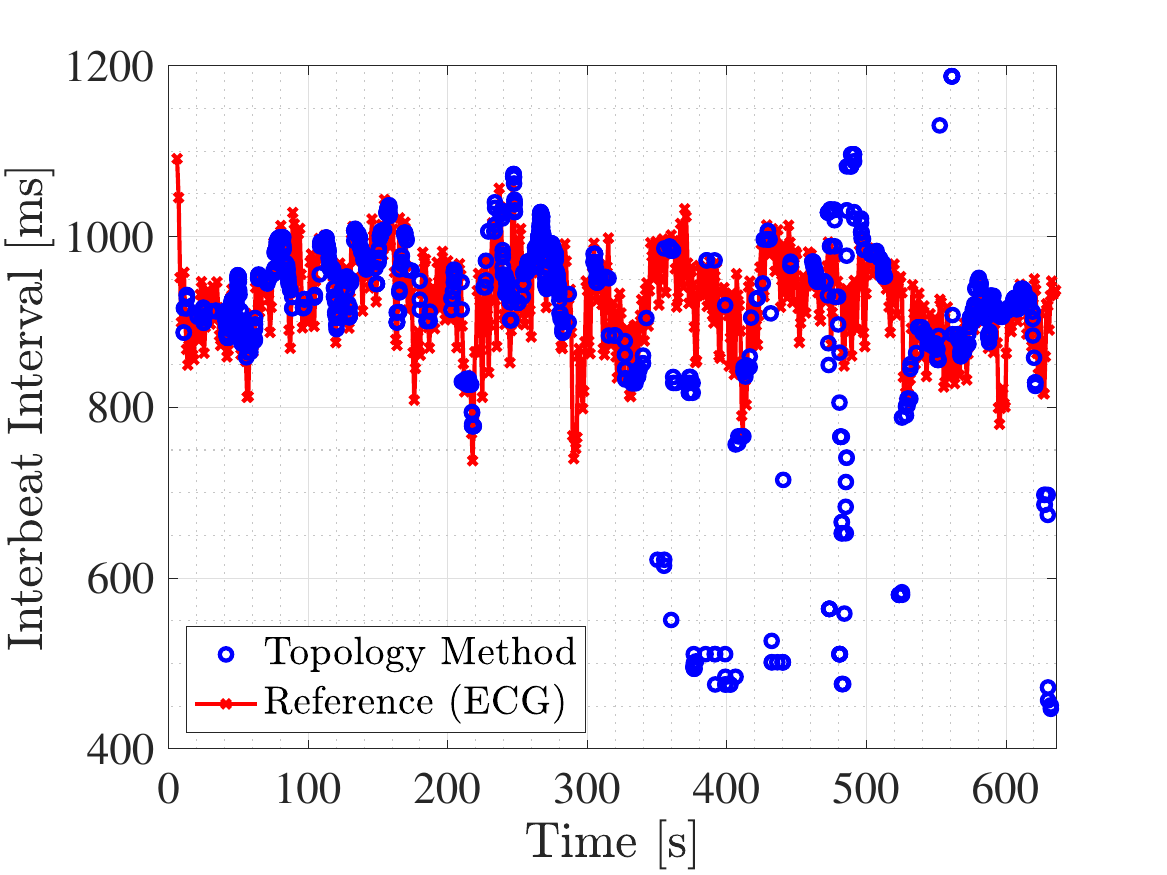}
    \subcaption{$\gamma = -0.5$}
    \label{fig:m2q}
  \end{minipage}
  \hspace{0.02\columnwidth} 
  \begin{minipage}[H]{0.32\linewidth}
    \centering
    \includegraphics[width=\columnwidth]{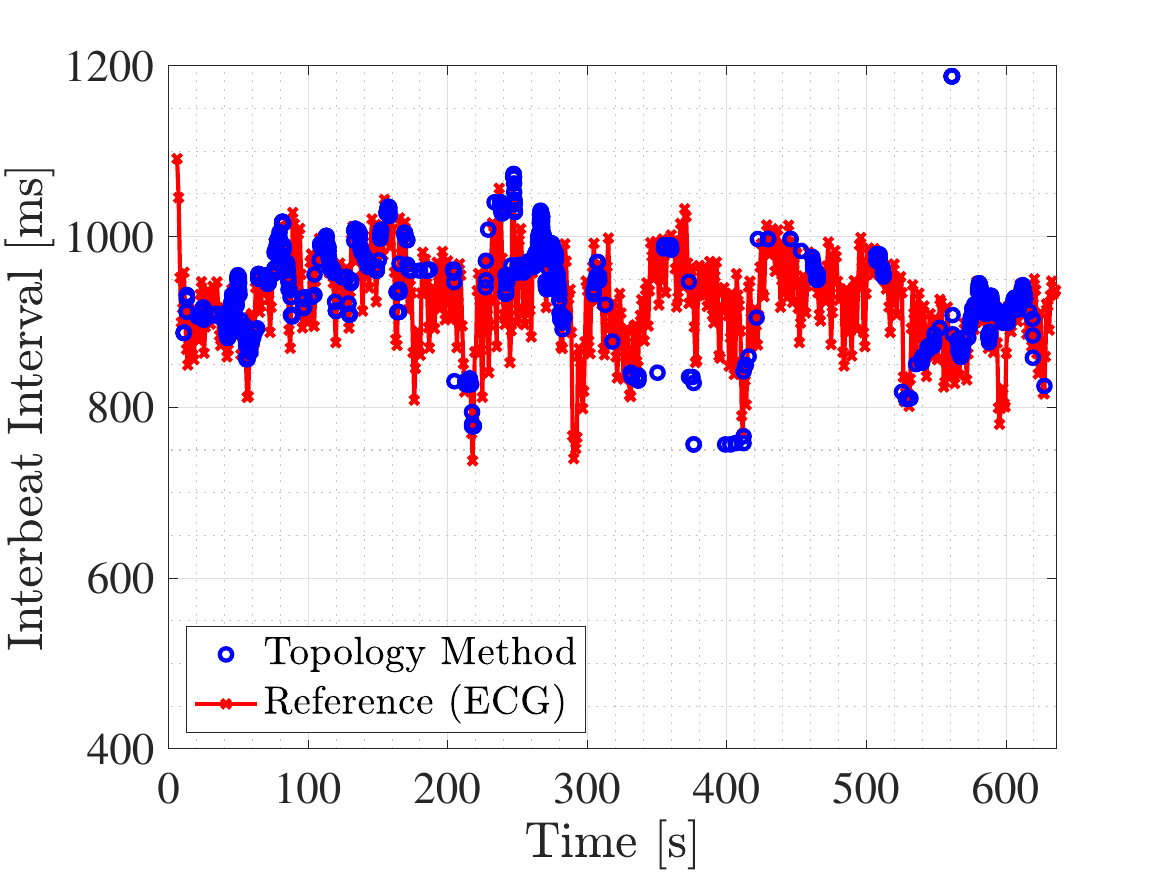}
    \subcaption{$\gamma = 0.625$}
    \label{fig:opt}
  \end{minipage}
  \caption{ Interbeat intervals (IBIs) estimated by applying the topology method (blue circles) to the data for subject 18 in the study of Schellenberger~et al.~\cite{bib:Schellenberger2020}. The red lines show the reference value for the IBI obtained from an electrocardiograph (The scaling factor $\gamma$ is defined in Fig.~\ref{fig:TopologyNum}(b)).} 
  \label{fig:topo18}
\end{figure*}

First, we show the accuracy and TCR of the topology method for participant 18 as an example.
Fig.~\ref{fig:topo18} shows the IBI estimated using the topology method  for $\gamma = 0$, $-0.5$, and $0.625$.
Note that $\gamma = -0.5$ corresponds to the ``simple assignment'' introduced in the original study~\cite{bib:Sakamoto2016}.
As shown in Fig.~\ref{fig:topo18}(a), the estimated values deviate more than 200 ms from the true value at times $370$~s~$\leq t \leq$~$410$~s and $460$~s~$\leq t \leq$~$490$~s for $\gamma = 0$, whereas Fig.~\ref{fig:topo18}(b) shows that the estimated values for $\gamma = -0.5$ deviate more than 200 ms from the reference value at times $350$~s $\leq t \leq 500$~s.
Comparing panels (a)\textendash (c) of Fig.~\ref{fig:topo18}, $\gamma=0.625$ appears to be the best value, but this must be confirmed using data from a cohort of participants.

\begin{figure}[tb]
	\centering
	\includegraphics[width=\linewidth]{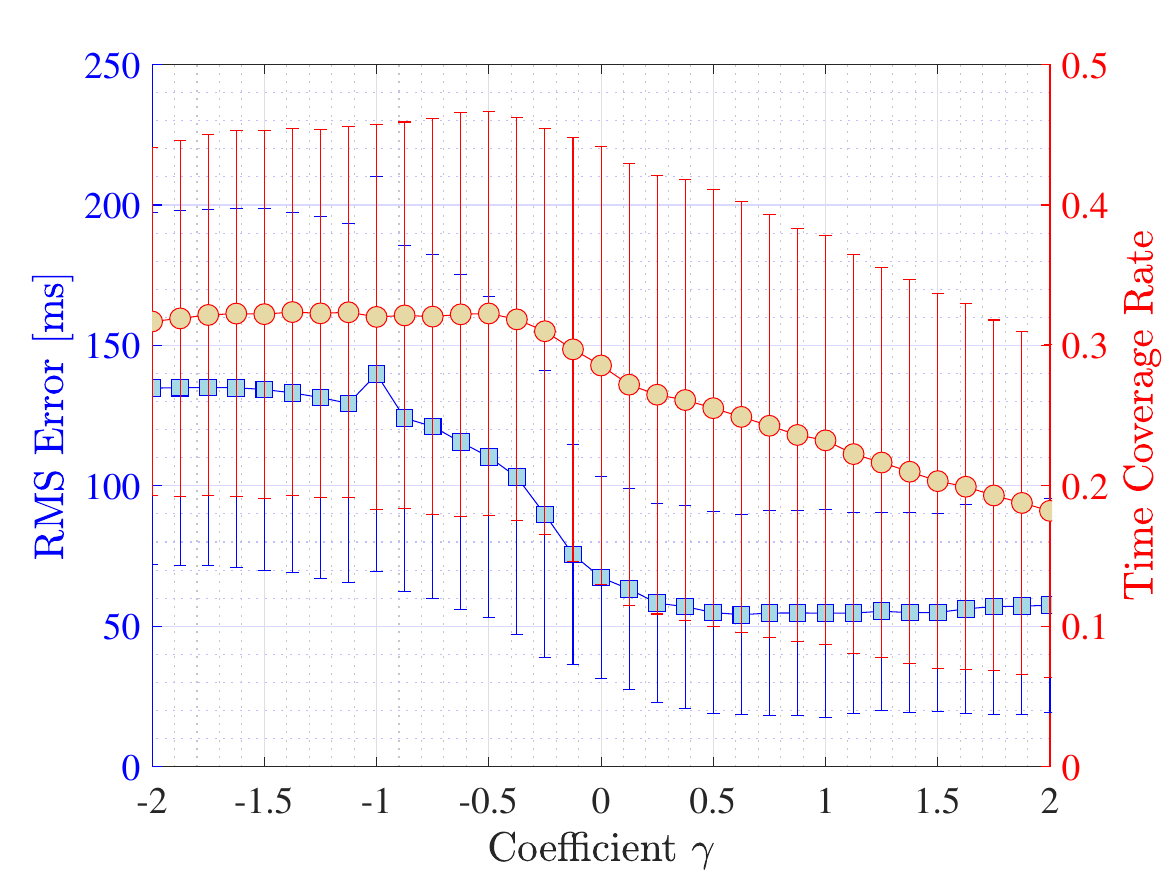}
	\caption{ Root mean square (RMS) error and time coverage rate (TCR) for different values of $\gamma$. Data show mean $\pm$ standard deviation ($n = 21$). }
	\label{fig:Topology_RMSEandTCR}
\end{figure}

Fig.~\ref{fig:Topology_RMSEandTCR} shows the RMS error and TCR of the topology method for different values of $\gamma$.
It can be seen that the RMS error for $\gamma \geq 0$ is smaller than that for $\gamma < 0$.
This result supports the validity of applying opposite signs to the values of RDP and RDV  (FDV and RDP), as shown in the simplified model analysis in the previous section.
As a function of $\gamma$, it was also confirmed that the RMS error decreases to a minimum that is reached at approximately $\gamma = 0.625$, whereas the TCR monotonically decreases as a function of $\gamma$ for $\gamma > 0$. 
These results indicate that $\gamma = 0.625$ is the optimal value for achieving both a low RMS error and high TCR.

From the above analysis, we confirmed that when $\gamma$ is approximately 0.625, the RMS error is smallest and results in minimal exclusion of data points from the IBI estimate.
Table~\ref{tbl:CmpRMSE} shows the comparison table of the RMS error.
We also evaluate the accuracy of the topology method when $\gamma$ takes a random number distributed uniformly between $-2$ and $2$; the RMS error was $85\pm40$ ms on average with a Monte Carlo simulation with 100 trials.
This finding supports the validity of the simplified model derived in section III. 
Notably, the RMS error and TCR obtained using $\gamma = 0.5$ are almost the same as those obtained using $\gamma = 0.625$.
This indicates the validity of the numbers $-\mathrm{j}/2$ and $\mathrm{j}/2$ assigned to RDV and FDP, respectively, in the original study~\cite{bib:Sakamoto2016}.

\begin{table}[tb]
\caption{Comparison table of the root mean square (RMS) error.} 
\begin{tabular}{c||c|c|c|c|}
 &  $\gamma = 0$ (a)  & $\gamma = -0.5$ (b) & $\gamma = 0.625$ (c) & $\gamma = 0.5$~\cite{bib:Sakamoto2016} \\
 \hline \hline
RMS error & $67 \pm 36$ $\mathrm{ms}$ & $110 \pm 57$ $\mathrm{ms}$ & $54 \pm 36$ $\mathrm{ms}$ & $55 \pm 36$ $\mathrm{ms}$\\
\hline
\end{tabular}
\label{tbl:CmpRMSE}
\end{table}

\section{Conclusion}
In this study, we examined the validity of complex number assignments to two types of inflection points, RDV and FDP, which are feature points in the topology method for estimating the IBI using millimeter-wave radar.
The validity was examined quantitatively using a simplified model of the skin displacement waveform and analysis of a publicly available dataset.
We derived the optimal assignments for RDV and FDP and demonstrated that these numbers yield an IBI estimate with a low RMS error and high TCR.
This suggests that although the complex numbers assigned to feature points by Sakamoto~ et al.~\cite{bib:Sakamoto2016} were empirical, they are near-optimal parameters for IBI estimation.
This study is the first to show the theoretical basis of the important parameter $\gamma$ in the topology method that has been applied to various radar-based heartbeat measurements.

\section*{Acknowledgment}
\addcontentsline{toc}{section}{Acknowledgment}
\scriptsize
This work was supported in part by SECOM Science and Technology Foundation, the Japan Science and Technology Agency (grant JPMJMI22J2), and the Japan Society for the Promotion of Science KAKENHI (grants 19H02155, 21H03427, 23H01420 and 23K19119). We thank Edanz (https://jp.edanz.com/ac) for editing a draft of this manuscript.

\end{document}